\documentclass{osa-article}

\journal{oe}
\articletype{Research Article}
\usepackage{lineno}
\renewcommand{\vv}[1]{\mathbf{#1}}
\newcommand{\ket}[1]{\bigl|#1\bigr>}
\newcommand{\bra}[1]{\bigl<#1\bigr|}
\newcommand{\braket}[2]{\bigl<#1\big|#2\bigr>}

\begin{document}

\title{Non-local polarization alignment and control in fiber using feedback from correlated measurements of entangled photons}

\author{
  Evan Dowling,\authormark{1,2},
  Mark Morris,\authormark{3},
  Gerald Baumgartner\authormark{3},
  Rajarshi Roy,\authormark{1,2,4},
  Thomas E. Murphy,\authormark{1,4,*}
}
\address{
  \authormark{1}Institute for Research in Electronics and Applied Physics, University of Maryland, College Park, Maryland 20740, USA\\
  \authormark{2}Department of Physics, University of Maryland, College Park, Maryland 20740, USA\\
  \authormark{3}Laboratory for Telecommunication Sciences, College Park, Maryland 20740, USA \\
  \authormark{4}Department of Electrical and Computer Engineering, University of Maryland, College Park, Maryland 20740, USA

  \email{\authormark{*}tem@umd.edu}
}

\begin{abstract}
  Quantum measurements that use the entangled photons' polarization to encode quantum information require calibration and alignment of the measurement bases between spatially separate observers.  Because of the changing birefringence in optical fibers arising from temperature fluctuations or external mechanical vibrations, the polarization state at the end of a fiber channel is unpredictable and time-varying. Polarization tracking and stabilization methods originally developed for classical optical communications cannot be applied to polarization-entangled photons, where the separately detected photons are statistically unpolarized, yet quantum mechanically correlated.  We report here a fast method for automatic alignment and dynamic tracking of the polarization measurement bases between spatially separated detectors.  The system uses the Nelder-Mead simplex method to minimize the observed coincidence rate between non-locally measured entangled photon pairs, without relying on classical wavelength-multiplexed pilot tones or temporally interleaved polarized photons. Alignment and control is demonstrated in a 7.1 km deployed fiber loop as well as in a controlled drifting scenario.
\end{abstract}

\section{Introduction}

Entanglement is a quantum mechanical phenomenon in which the outcomes of spatially separated measurements are dependent in a way that cannot be explained classically \cite{Nielsen2012}.  Entangled states are essential ingredients in versions of quantum key distribution, quantum communication, quantum teleportation and quantum computing.  The control and characterization of entanglement is essential in all of these applications \cite{Nielsen2012,Giustina2015,Bennett1992,Jin2015}.

Entanglement can be arranged between atoms, electrons, and even larger, more complex physical systems, but the most convenient and practical method to transmit entangled states is through photons, which can be conveyed over low-loss optical optical fibers \cite{Schultz1979,Miya1979}, allowing far greater spatial separation of the entangled measurements \cite{Dynes2009,Inagaki2013}. One of the most prevalent methods to generate entangled photons is through spontaneous parametric downconversion (SPDC) -– a nonlinear optical process in which a single pump photon spontaneously splits into a pair of lower-energy emitted photons \cite{Hong1985,Burnham1970,Zhang2021}.  While other types exist, in type II spontaneous parametric downconversion, the two generated photons are orthogonally polarized.  When the SPDC process is tuned to degeneracy, the emitted photons are spectrally indistinguishable, but orthogonally polarized.  When these photons are split and transmitted to separate observers, if the polarization state of one photon is measured, the result is completely indeterminate but anticorrelated with the observed polarization state of the other photon.

A significant obstacle for entanglement distribution through fiber is that the majority of deployed fiber communication channels do not preserve the state of polarization of the optical signal.  Even if the input polarization state is known, the state that emerges from a span of single-mode fiber is indeterminate and varies in time because of unpredictable and uncontrolled variations in temperature, bending, and stress \cite{Imai1988,Namihira1989,Karlsson2000,Waddy2005,Ding2017a}.  Observation of polarization entanglement relies upon alignment and calibration of the measurement bases between the two separate observers, which is hindered by these uncontrolled variations.

The problem of polarization alignment and stabilization is well-studied in classical fiber transmission systems\cite{Noe1988,Walker1990,Shimizu1991,Heismann1994,Martinelli2003,Martinelli2006}, but existing methods cannot be directly applied when the measured signals are at the single-photon level.  Moreover, if the photons are polarization-entangled, the received photons in each fiber are statistically unpolarized, and polarization control cannot be achieved from local measurements alone.

One solution to this problem is to interleave well-polarized alignment signals, either in time or wavelength, that co-propagate with the quantum signal in the same fiber \cite{Xavier2008,Xavier2009,Chen2009,Li2018,Wang2009,Chung2022}. While the injection of classical alignment signals, interleaved in time or wavelength, is a well-established strategy that may offer faster alignment than the method described here, there are also drawbacks that must be weighed, depending on the application.  Time multiplexing can limit the bandwidth of the transmitted quantum signal because of leakage of the pilot tone into the quantum signal when repetition rates are too high. Wavelength multiplexing places similarly stringent constraints on optical filtering, and is ultimately limited by polarization mode dispersion between the signal and pilot wavelengths.  Both methods add experimental complexity as they require pilot tone sources, multiplexers, and detectors for integration, separation, and measurement of these classical signals along with the quantum sources and detectors.

In QKD systems that employ the BB84 protocol, the transmitted photons are randomly polarized, but not entangled.  In this case, the polarization axes of the receiver can be aligned locally by using only the sifted keys or basis reconciliation process, which requires only single-photon measurements\cite{Ding2017b,Agnesi2020}. More recently, Shi {\it et al.} achieved polarization alignment in an entanglement-based QKD system, by using a stochastic optimization algorithm to minimize the quantum bit error rate between non-local receivers\cite{Shi2021}.  They report compensation in under 20 minutes, which is sufficient for tracking slower polarization drifts in shorter spans of underground deployed fiber. In addition, an alternative polarization alignment strategy using Bayesian estimation was shown to work in low coincidence count, high shot noise regimes \cite{Cortes2022}.

We report here an experimental method for both polarization alignment and tracking that uses a simplex optimization algorithm based on non-local coincidence measurements to acquire and stabilize the observation of polarization-entangled photons.  The method requires no temporal or spectral interleaving of classically polarized signals, and we confirm that when measured locally, the received photons carry no signature of the polarization state.  Experimental results show that the method achieves alignment within approximately 160 seconds, to target states placed arbitrarily on the Poincar\'e sphere.  The system is shown to dynamically adapt to systematic variations in the opposite (uncontrolled) fiber channel.  We finally demonstrate the successful operation of the system over a 7.1 km deployed fiber link.  This method could find applications in key distribution, quantum networking, quantum teleportation, and quantum measurement.

\section{Experiment}

\begin{figure}[htbp]
    \centering
    \includegraphics[]{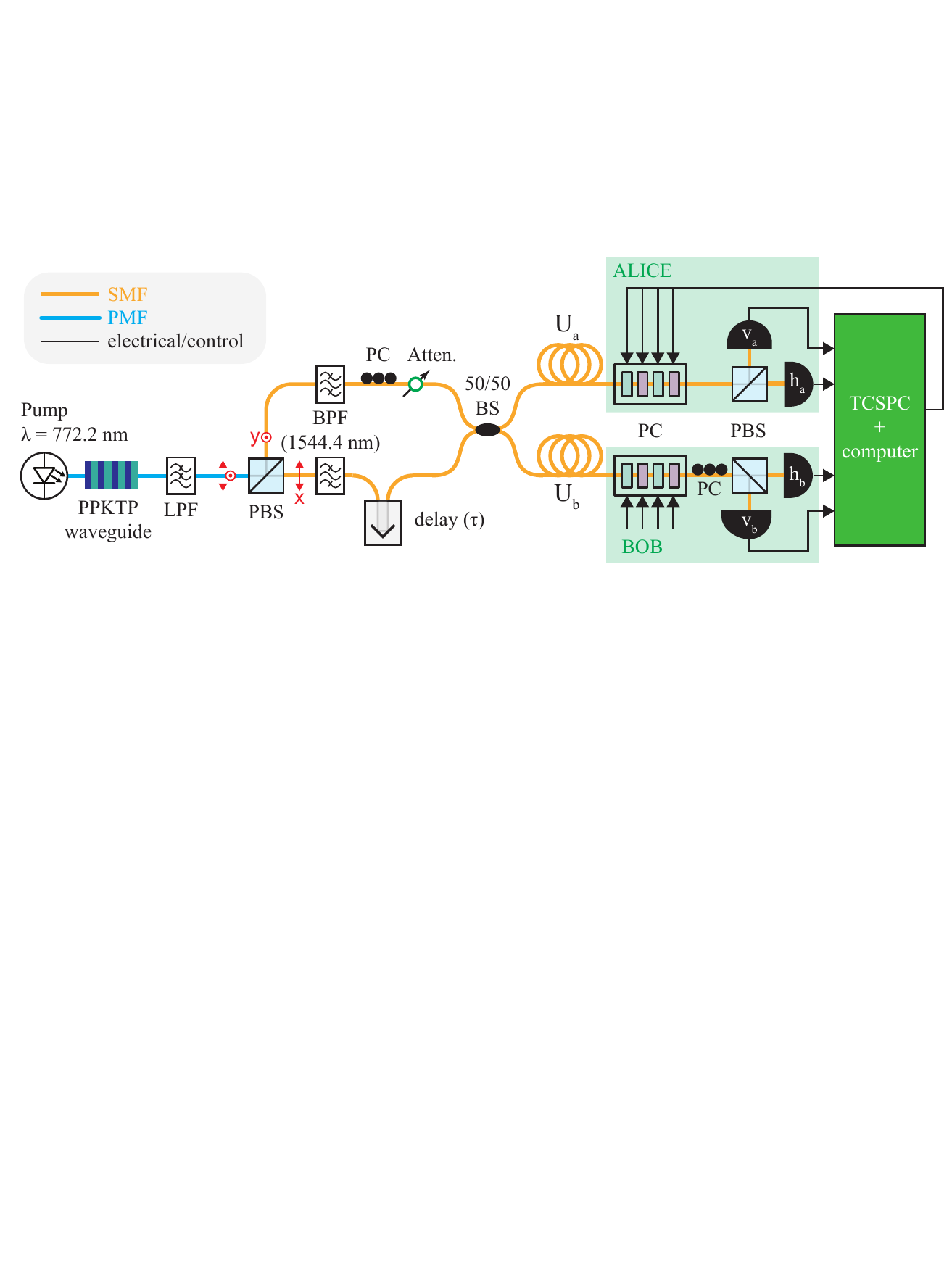}
    \caption{Diagram of experimental setup used to demonstrate non-local polarization control.  Entangled photons are generated through spontaneous parametric downconversion in a periodically poled potassium titanyl phosphate (PPKTP) waveguide, followed by longpass and bandpass filters (LPF, BPF) to extinguish the pump signal and ensure spectral indistinguishability (Fig.~\ref{fig:2}).  The photons are transmitted to spatially separated receivers (Alice and Bob), which each record the photons after polarizing beamsplitters (PBS).  Piezoelectric polarization controllers (PC) adjust the polarization state in each channel. A time-correlating single-photon counting (TCSPC) instrument records the coincidence rate. }
    \label{fig:1}
\end{figure}

Fig.~\ref{fig:1} shows the experimental apparatus used to demonstrate non-local polarization alignment and control.  Polarization-entangled photons are generated using Type II spontaneous parametric downconversion in a periodically-poled potassium titanyl phosphate (PPKTP) waveguide (AdvR, Inc.).  The PPKTP waveguide is pumped with a 772.2 nm, continuous-wave external cavity diode laser (Newport TLB-7100), which generates $\sim 1\times 10^6$ pairs of orthogonally-polarized, wavelength degenerate photons per second\cite{Fiorentino2007,Couteau2018}.  An anti-reflection coated silicon wafer and 1550 nm long pass filter are used to extinguish any residual pump light and other noise. Because of the birefringence of both the PPKTP waveguide and the polarization maintaining (PM) output fiber, the downconverted photons acquire a differential group delay, making them distinguishable\cite{Kuklewicz2004}.  The downconverted photons are separated in a polarizing beamsplitter, with a controllable time delay stage in one arm to equalize the timing.

\begin{figure}[htbp]
    \centering
    \includegraphics[]{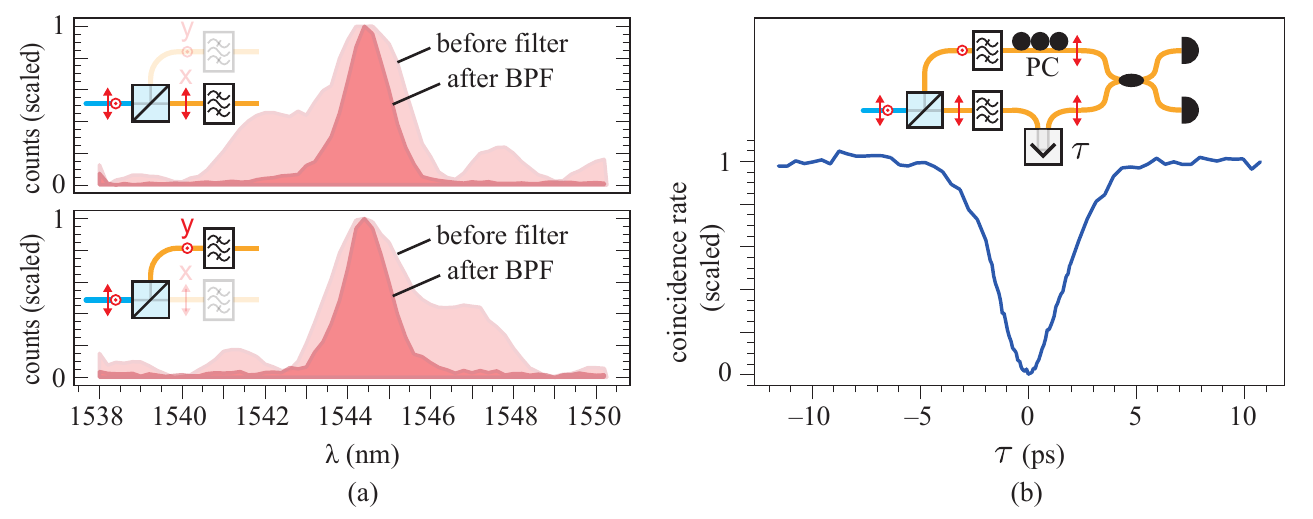}
    \caption{a) Spontaneous parametric downconversion spectrum from PPKTP waveguide before and after the bandpass filters (BPF), showing down converted signal (H polarized) and idler (V polarized) photons. b) Measured Hong-Ou-Mandel interference, demonstrating indistinguishability when the two combined, filtered photons are co-polarized and coincident.}
    \label{fig:2}
\end{figure}

Fig.~\ref{fig:2}a shows the spectra of the spontaneously-generated $x$- and $y$-polarized photons, measured after the polarization beamsplitter.  A pair of matching 1.2 nm tunable bandpass filters (OZ Optics) were used to extinguish the distinguishable spectral tails caused by asymmetry in the phase matching conditions.  The time delay ($\tau$) was adjusted by co-polarizing and combining the channels and recording the Hong–Ou–Mandel (HOM) interference, which shows 96\% extinction in the coincidence rate, confirming the indistinguishability of the photons, as shown in Fig.~\ref{fig:2}(b).

To generate entangled photons, the polarization controllers in the arms of the source interferometer are then adjusted so that the two photons arrive at the (non-polarizing) 50:50 beam splitter with orthogonal polarizations.  A variable attenuator is used to equalize the powers in the two polarization states, in order to eliminate any partial polarization component for the transmitted photons.  After the beam splitter the entangled photons are then transmitted, optionally through a fiber span, to the receivers.

The two spatially separated receivers (here denoted Alice and Bob) each include a piezoelectric fiber-based polarization controller (General Photonics PolaRITE III), polarization beamsplitter and a pair of superconducting nanowire photon-counting detectors with quantum efficiencies of 80\%  [Opus One, Quantum Opus LLC.].  Because the nanowire detectors have a polarization-dependent efficiency, an additional manual polarization controller (not shown) is inserted between the polarizing beamsplitter output and the photon counter to maximize the quantum efficiency.  The electronic signals are time-tagged and correlation measurements are processed in real time by a time-correlating single-photon counting (TCSPC) instrument [HydraHarp 400, PicoQuant LLC.] with timing resolution up to 1 ps, which enables measurement of the pairwise coincidence rate between spatially separate detectors.  Due to detector and electronic jitter a coincidence window of 100 ps is used for signal detection. By using a small coincidence window, we improve the signal-to-noise ratio by ignoring the unpaired detection events that do not temporally coincide with our downconverted photons. A computer records the coincidence rate in real time and uses the resulting measurement to control one or both of the piezoelectric polarization controllers.


\section{Theory}

The output entangled photon pair state emerging from the beam splitter can be expressed as:
\begin{equation}\label{eq:1}
  \ket{\Psi_{\rm out}} = \frac{1}{2}\left( \ket{x_a}\ket{y_b} - \ket{y_a}\ket{x_b} + \ket{x_a}\ket{y_a} + \ket{x_b}\ket{y_b}\right)
\end{equation}
where, for example, the leading term $\ket{x_a}\ket{y_b}$ describes an $x$-polarized photon transmitted to Alice's port and a $y$-polarized photon transmitted to Bob's port.  The first two terms in \eqref{eq:1} represent the case when pairwise orthogonally polarized photons are transmitted to Alice and Bob, respectively, while the final terms represent the possibility that both photons go to either Alice or Bob.  Because our experimental measurements are conditioned on coincidences between Alice and Bob, these final terms are excluded through post-selection\cite{Kuklewicz2004,Takeoka2015,Brewster2021}.  The post-selected state then simplifies to the $\ket{\Psi^-}$ Bell state,
\begin{equation}\label{eq:2}
  \ket{\Psi^-} = \frac{1}{\sqrt{2}}\left(\ket{x_a}\ket{y_b} - \ket{y_a}\ket{x_b} \right)
\end{equation}

The fiber connecting to Alice transforms the polarization state of the optical signal, so that an initially $x$-polarized state will, in general, emerge from the fiber as elliptically polarized, and conversely, an emerging linear polarization state would correspond to an elliptically-polarized launch state.  This transformation between the input and output states can be described by a complex unitary matrix (Jones matrix) $\vv{U}_a$ and its inverse:
\begin{equation}\label{eq:3}
  \begin{bmatrix}
    \ket{h_a} \\
    \ket{v_a} \\
  \end{bmatrix} =
  \vv{U}_a
  \begin{bmatrix}
    \ket{x_a} \\
    \ket{y_a} \\
  \end{bmatrix},\qquad
  \begin{bmatrix}
    \ket{x_a} \\
    \ket{y_a} \\
  \end{bmatrix} =
  \vv{U}_a^\dagger
  \begin{bmatrix}
    \ket{h_a} \\
    \ket{v_a} \\
  \end{bmatrix}
\end{equation}
where $\ket{h_a}$ and $\ket{v_a}$ denote the two output states that get separated and detected by the polarizing beamsplitter and photon detectors.  The matrix $\vv{U}_a^\dagger$ describes how these output states are mapped back into an equivalent elliptical polarization basis at Alice's fiber input.  A general form for a unitary matrix (ignoring the common phase associated with transmission) is:
\begin{equation}\label{eq:4}
  \vv{U}_a^\dagger =
    \begin{bmatrix}
      a_x & -a_y^* \\
      a_y & a_x^* \\
    \end{bmatrix}
\end{equation}
where the first column of $\vv{U}_a^\dagger$ is a complex unit vector (Jones vector),
\begin{equation}\label{eq:5}
  \vv{\tilde a} \equiv
    \begin{bmatrix}
      a_x \\
      a_y \\
    \end{bmatrix},\qquad |a_x|^2 + |a_y|^2 = 1\quad.
\end{equation}
The vector $\vv{\tilde a}$ defines the input polarization state that would be fully transmitted into the ``h'' output of Alice's polarization beamsplitter.  A polarizing beamsplitter that is preceded by a unitary transformation is thus equivalent to an elliptical polarization beamsplitter, that splits the incident light along an axis defined by the Jones vector $\vv{\tilde a}$.

The complex, two-dimensional Jones vector $\vv{\tilde a}$ can be translated into a real, three-dimensional Stokes vector $\vv{a}$,
\begin{equation}\label{eq:6}
  \vv{a} =
    \begin{bmatrix}
      a_1 \\
      a_2 \\
      a_3 \\
    \end{bmatrix} =
    \begin{bmatrix}
      |a_x|^2 - |a_y|^2 \\
      a_x a_y^* + a_x^* a_y \\
      i(a_x a_y^* - a_x^* a_y) \\
    \end{bmatrix},\quad a_1^2 + a_2^2 + a_3^2 = 1
\end{equation}
that is easily visualized as a vector on the Poincar\'e sphere.  $\vv{a}$, and its temporal variations can also be measured experimentally by transmitting a classical linearly polarized signal through the fiber and recording the output state on a polarimeter.  Bob's channel undergoes a similar, but different polarization transformation, described by analogous Jones and Stokes vectors $\vv{\tilde b}$ and $\vv{b}$ that characterize the polarization rotation occurring in Bob's fiber.

\begin{figure}[htbp]
    \centering
    \includegraphics{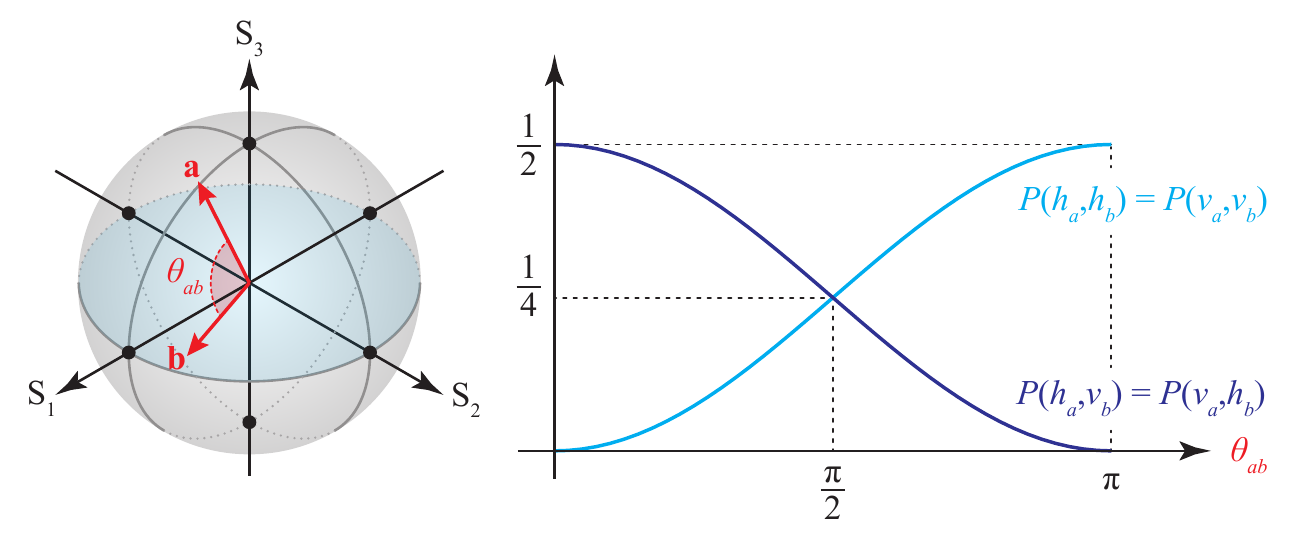}
    \caption{Alice's and Bob's horizontal measurement Stokes vectors, $\vv{a}$ and $\vv{b}$ on the Poincar\'e sphere. The joint probability of detection from the shared entangled state \eqref{eq:2} is expressed in equation \eqref{eq:10} as a function of the interior angle $\theta_{ab}$.}
    \label{fig:3}
\end{figure}

With these definitions, the probability that Alice and Bob both detect horizontally polarized photons can then be calculated as
\begin{equation}\label{eq:7}
  P(h_a,h_b) = \left|\bra{h_a} \braket{h_b}{\Psi^-} \right|^2
\end{equation}
where we use \eqref{eq:3}-\eqref{eq:4} to express
\[
  \ket{h_a} = a_x^*\ket{x_a} +  a_y^*\ket{y_a},\quad \ket{h_b} = b_x^*\ket{x_b} +  b_y^*\ket{y_b}
\]
Upon substitution into \eqref{eq:7}, this yields, after algebraic simplification,
\begin{equation}\label{eq:8}
  P(h_a,h_b) = \frac{1}{2}\left| a_x b_y - a_y b_x  \right|^2 = \frac{1}{2}\left(1 - \left|\vv{\tilde a}^*\cdot\vv{\tilde b}\right|^2\right)
\end{equation}
The inner product of two Jones vectors $\vv{\tilde a}^*\cdot\vv{\tilde b}$ can be related to the inner product of their corresponding Stokes vectors\cite{Gordon2000},
\begin{equation}\label{eq:9}
  \left|\vv{\tilde a}^*\cdot\vv{\tilde b}\right|^2 = \frac{1}{2}\left(1+\vv{a}\cdot\vv{b}\right)
\end{equation}
which gives a simple expression for the coincidence probability
\begin{equation}\label{eq:10}
  P(h_a,h_b) = \frac{1}{4}\left(1 - \vv{a}\cdot\vv{b}\right) = \frac{1}{4}(1-\cos\theta_{ab})
\end{equation}
where $\theta_{ab}$ represents the angular separation between polarization bases $\vv{a}$ and $\vv{b}$ on the Poincar\'e sphere, as shown in Fig.~\ref{fig:3}.  The coincidence rate between horizontal and vertical channels, is, analogously, found to be:
\begin{equation}\label{eq:11}
  P(h_a,v_b) = \frac{1}{4}\left(1 + \vv{a}\cdot\vv{b}\right) = \frac{1}{4}(1+\cos\theta_{ab})
\end{equation}
These simple relationships are a generalization of the well-known result that for linearly polarized measurements of entangled photons, the coincidence rate depends only on the relative difference between their polarization axes.  When $\vv{a}$ and $\vv{b}$ are matched, the coincidence rate is zero, as expected.  When $\vv{a}$ and $\vv{b}$ are anti-aligned on the Poincar\'e sphere (which corresponds to orthogonal bases), the coincidence rate is 1/2, because the photons may emerge from the other pair of orthogonal output channels with equal probability.

An important consequence of equation \eqref{eq:10} is that while non-local measurements are required to measure the coincidence rate, the adjustment needed to minimize the coincidence rate only needs to be performed at one location, i.e., Alice may adjust $\vv{a}$ to any target state $\vv{b}$ in order to minimize \eqref{eq:10}.

This simple unitary matrix transformation describes a static, spectrally uniform polarization transformation, and ignores fiber impairments like chromatic dispersion, polarization mode dispersion, and polarization-dependent loss.  These effects become increasingly significant for longer fiber spans, or if the photons occupy a broader spectral bandwidth.  If left uncompensated or imbalanced between the channels, transmission imperfections will alter the density matrix of the two-photon state and impair the observation of coincidences between the channels\cite{Jones2018}.

The polarization adjustment is achieved using piezoelectric fiber polarization actuators\cite{Shimizu1991}, which offer electric polarization control with negligible fiber insertion loss, an important advantage over liquid crystal actuators, electrooptic actuators, or rotating waveplates.  The piezoelectric actuators provide four successive squeezing elements that produce a strain-induced fiber birefringence oriented at $0^\circ$, $45^\circ$, $0^\circ$, and $45^\circ$.  Each actuator produces a retardance in proportion to the applied voltage, which can be visualized on the Poincare sphere as an adjustable rotation about the axes $S_1$ or $S_2$ (in the reference frame of the piezoelectric actuator.)

In principle, if Alice's PBS is adjusted to be $45^\circ$ relative to axis of the final piezoelectric actuator (i.e., if the PBS is oriented along $S_1$), then it would be possible to achieve complete alignment of $\vv{a}$ to $\vv{b}$ by using only the two piezoelectric stages of the actuator \cite{Martinelli2003}.  In practice, however, this condition is difficult to achieve.  We therefore use three piezoelectric stages, which ensures full coverage of the Poincar\'e sphere regardless of how the PBS is aligned.  The fourth actuator could be employed to achieve reset-free actuation\cite{Noe1988}, but was not used here.

We chose to employ the Nelder-Mead simplex algorithm \cite{Nelder1965} to minimize the measured rate of coincidence detection, by adjusting the piezoelectric polarization controller at Alice. The algorithm creates a four-vertex tetrahedral simplex in the 3D space defined by the voltages on three actuators.  The Nelder-Mead method is agnostic to the actuation mechanism, and does not rely on a mathematically-predictive model to determine the actuation voltages. Because it iteratively refines the actuator voltages in order to minimize the coincidence rate, it can be executed continuously in order to track sufficiently slow changes in the minimization condition.  The Nelder-Mead algorithm is implemented through python’s scipy library \cite{Virtanen2020}.

\section{Polarization Alignment}

\begin{figure}[htbp]
    \centering
    \includegraphics[]{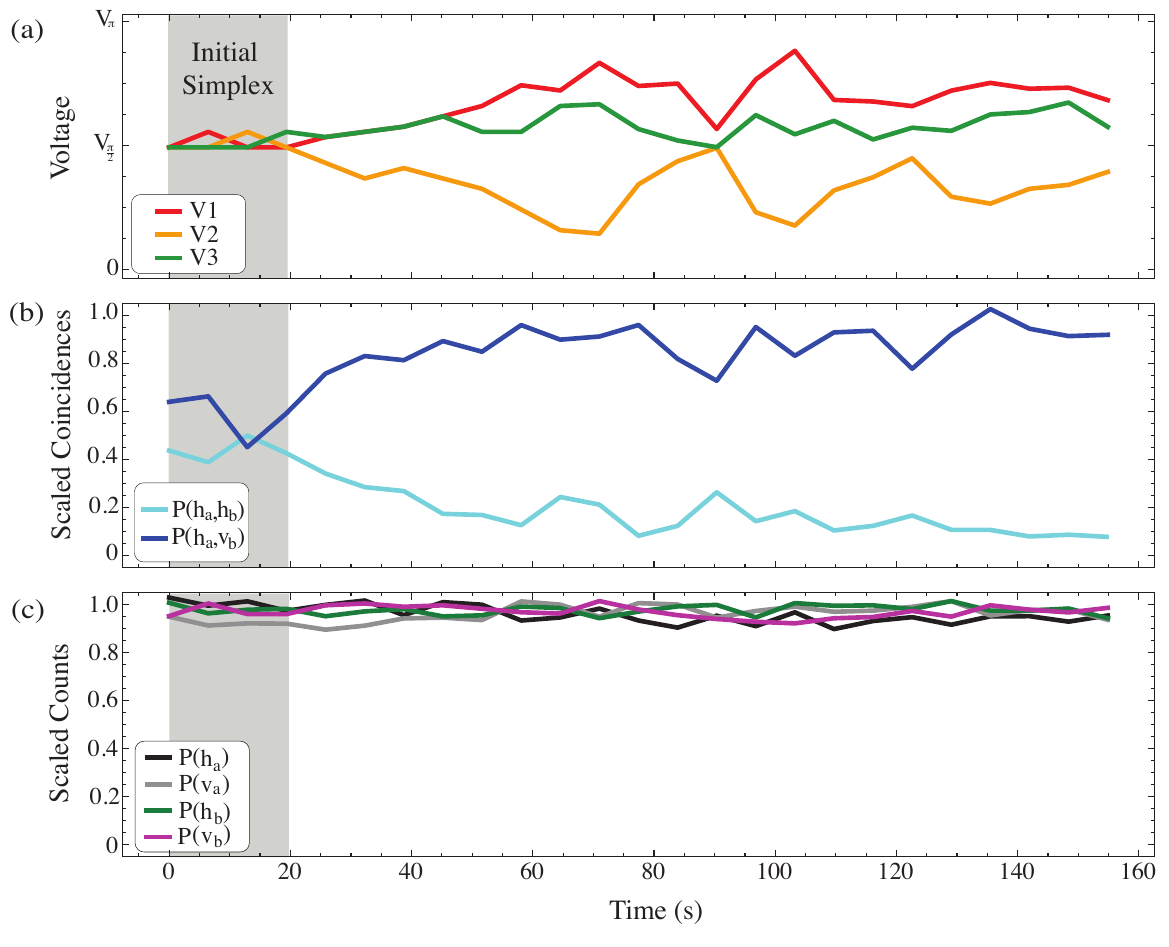}
    \caption{Representative example showing the adaptive alignment of Alice's measurement basis to minimize the observed coincidence rate.  (a) The three voltages applied to Alice's piezoelectric polarization controller, which change as they iterate through the Nelder-Mead algorithm.  (b) The Nelder-Mead algorithm finds the minimum of the measured $h_a$ and $h_b$ photon coincidence rate, seen in cyan. Conversely, the joint photodetection between the $h_a$ and $v_b$ photons rises to a maximum, seen in blue.  (c) The single photon count rates remain unchanged for both Bob and Alice throughout the alignment process, indicating no polarization dependence within those signals. }
    \label{fig:4}
\end{figure}

An example of the convergence between Alice's and Bob's horizontal measurement bases can be seen in Fig.~\ref{fig:4}. Alice's three voltages, Fig.~\ref{fig:4}a, change as they iterate through the Nelder-Mead algorithm minimizing the coincidence rate between Alice's and Bob's horizontal detectors, cyan line in Fig.~\ref{fig:4}b. The initial simplex begins around half the half-wave voltage for each piezoelectric actuator, denoted $V_\pi$. After thirty measurements the extinction ratio between the coincidences of the two detectors being minimized ($h_a, h_b$) and the coincidences of the detectors being maximized ($h_a, v_b$) raises above 90\%. It is important to note that Bob's and Alice's count rates, Fig.~\ref{fig:4}c, remain constant even though Alice's polarization controller is changing and Bob's remains fixed. This observation confirms that the alignment is reliant on the non-local entangled measurement, and that the local single photon signals do not carry any polarization signature.

\begin{figure}[htbp]
    \centering
    \includegraphics[]{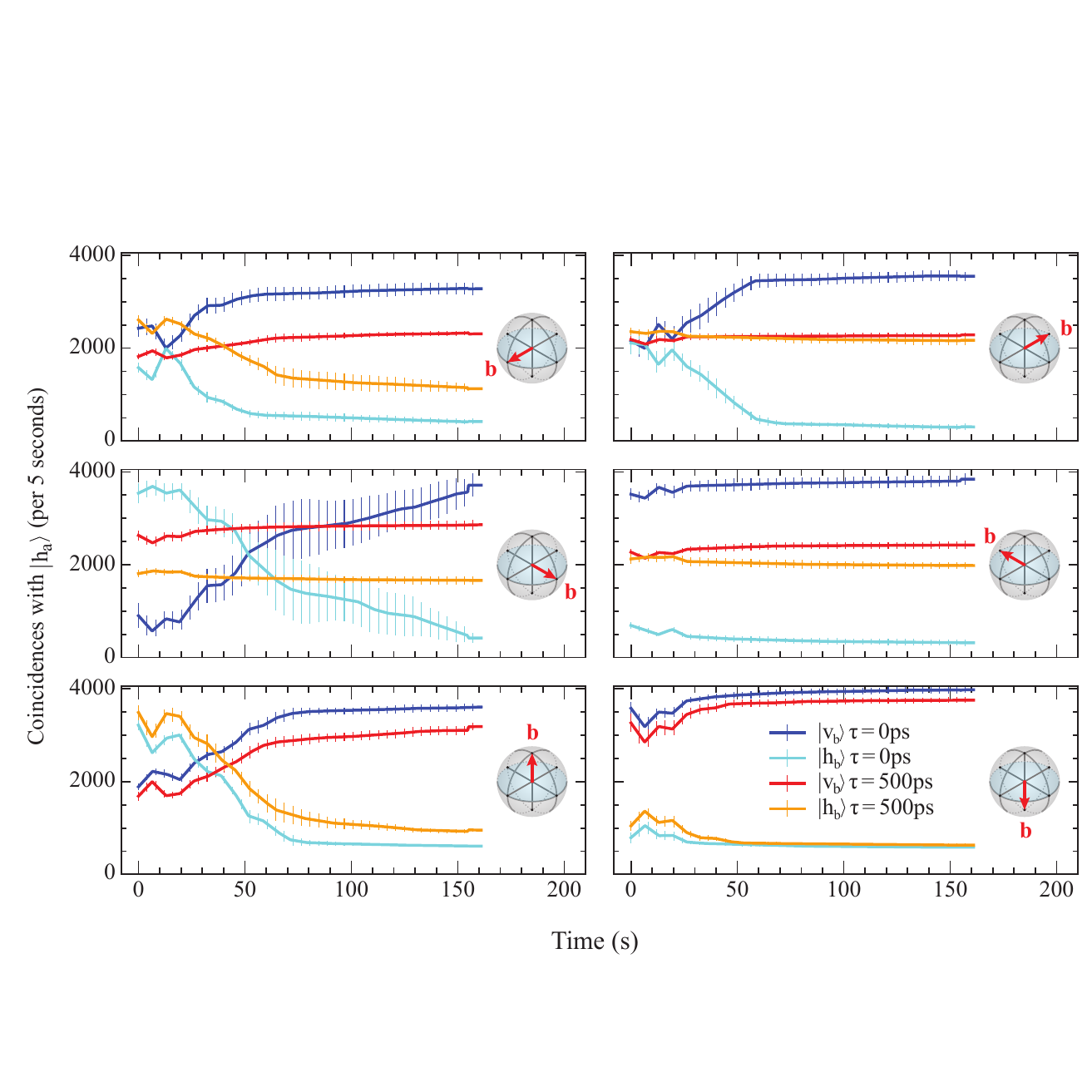}
    \caption{Automatic alignment of Alice's polarization basis, observed for six different target states of Bob's measurement basis ($\vv{b}$), distributed as the six poles on the Poincar\'e sphere.  In all cases, Alice's polarization controller starts at $\vv{V} = [V_\pi/2,V_\pi/2,V_\pi/2]$.  Time averaging and standard deviations are shown from 20 identical runs for each configuration. Coincidence counts for Bob's two detectors ($h_b, v_b$) with Alice's ($h_a$) detector for both entangled ($\tau = 0$ ps) and non-entangled ($\tau=500$ ps) photons are shown in cyan, blue, orange, and red respectively. Each initial configuration of Bob's $h_b$ measurement direction is shown to the right of every panel. }
    \label{fig:5}
\end{figure}

To assess the robustness of our alignment algorithm, we used a second piezoelectric actuator in Bob's channel to manually set the target basis ($\vv{b}$), and then allowed the alignment algorithm to automatically adjust Alice's basis ($\vv{a}$) to match.  Bob's measurement basis was pre-established by back-propagating a classical 1550 nm CW laser through one of the PBS exit ports and recording the resulting polarization state on a polarimeter inserted at the 50/50 beamsplitter.  Bob's piezoelectric voltages were then manually adjusted (and recorded) in order to map his polarization basis to the six nodes of the Poincar\'e sphere, $\pm S_1, \pm S_2, \pm S_3$, as shown in the insets of Fig.~\ref{fig:5}. After re-connecting the entangled photon source and photon counters, we cycled through these six pre-configured target states, each time starting Alice at the same initial condition of $\vv{V}_{\rm Alice} = [V_\pi/2,V_\pi/2,V_\pi/2]$.

The cyan traces in Fig.~\ref{fig:5} plot the measured coincidence rate between Alice's $h_a$ photon and Bob's $h_b$ photon, showing successful convergence with an extinction ratio of 95\% within 160 seconds in all cases.  The error bars indicate the statistical spread obtained from 20 independent trials.  The dark blue traces show the complementary coincidence rate between $h_a$ and $v_b$ photons, which is maximized when the $P(h_a,h_b)$ is minimized.

The algorithm assumes polarization entanglement between the two detected photons, which means that while coincidences are measured non-locally, the relative alignment and tracking can be actuated entirely at one observer (Alice), even if the opposite channel is drifting or changing.  We therefore expect this relative alignment method to fail when this entanglement is impaired or destroyed.  To test this, we used the variable delay line to introduced a 500 ps temporal delay between the signal and idler photons prior to the 50/50 beamsplitter, which destroys the polarization entanglement.  The orange (and red) traces in Fig.~\ref{fig:5} show that in this case, the algorithm is unable to consistently minimize (and maximize) the coincidence rate between Alice and Bob, except for the case of $\vv{b} = \pm S_3$, which we presume was coincidentally close to the case of Bob's axes being aligned to the signal and idler axes of the PPKTP nonlinear crystal.

Quantum measurements often require coordinated observation of coincidence rates along multiple, non-orthogonal directions.  Although the method presented here successfully aligns the polarization bases of two spatially separate observers, it does not simultaneously align the non-orthogonal polarization axes.  While simultaneous alignment could be achieved by adding additional polarization beamsplitters and detectors, or by multiplexing classical pilot tones in wavelength or time\cite{Xavier2008}, a more common and economical approach is to sequentially cycle through different detection bases, which could be pre-determined using the alignment method presented here, as demonstrated in Fig.~\ref{fig:5}.

\section{Alignment Accuracy}

Simulations of our experiment allow us to examine how experimental parameters affect the Nelder-Mead algorithm's ability to align Alice's and Bob's measurement bases. Under stationary conditions, the degree of polarization alignment and attainable extinction ratio in the coincidence rates depends on the integration period per measurement, $T$, and two counting rates, $r_p$ representing the maximum rate of pair detection (when $\theta_{ab}=\pi$), and $r_a$ the background accidental coincidence rate. In experiments and simulations, the algorithm seeks to minimize the total measured coincidence count,
\begin{equation}
    N = N_p+N_a
\end{equation}
where, using the result from \eqref{eq:10}, we model $N_p$ as a Poisson-distributed random variable with mean $r_p T\left(1-\cos{\theta_{ab}}\right)$ and $N_a$ as an independent Poisson-distributed random variables with mean $r_a T$. Taken together, the number of coincidences, $N$, measured in an integration interval, $T$, can be modeled by a discrete Poisson distribution with a mean that depends on the misalignment angle $\theta_{ab}$
\begin{equation}
    p_N(k) = P(N=k) = \frac{\bigl(r(\theta_{ab})T\bigr)^k}{k!}e^{-r(\theta_{ab})T}\ ,  \qquad  r(\theta_{ab}) = r_p\left(1-\cos{\theta_{ab}}\right) + r_a \quad .
    \label{eq:13}
\end{equation}

\begin{figure}
    \centering
    \includegraphics[width=\textwidth]{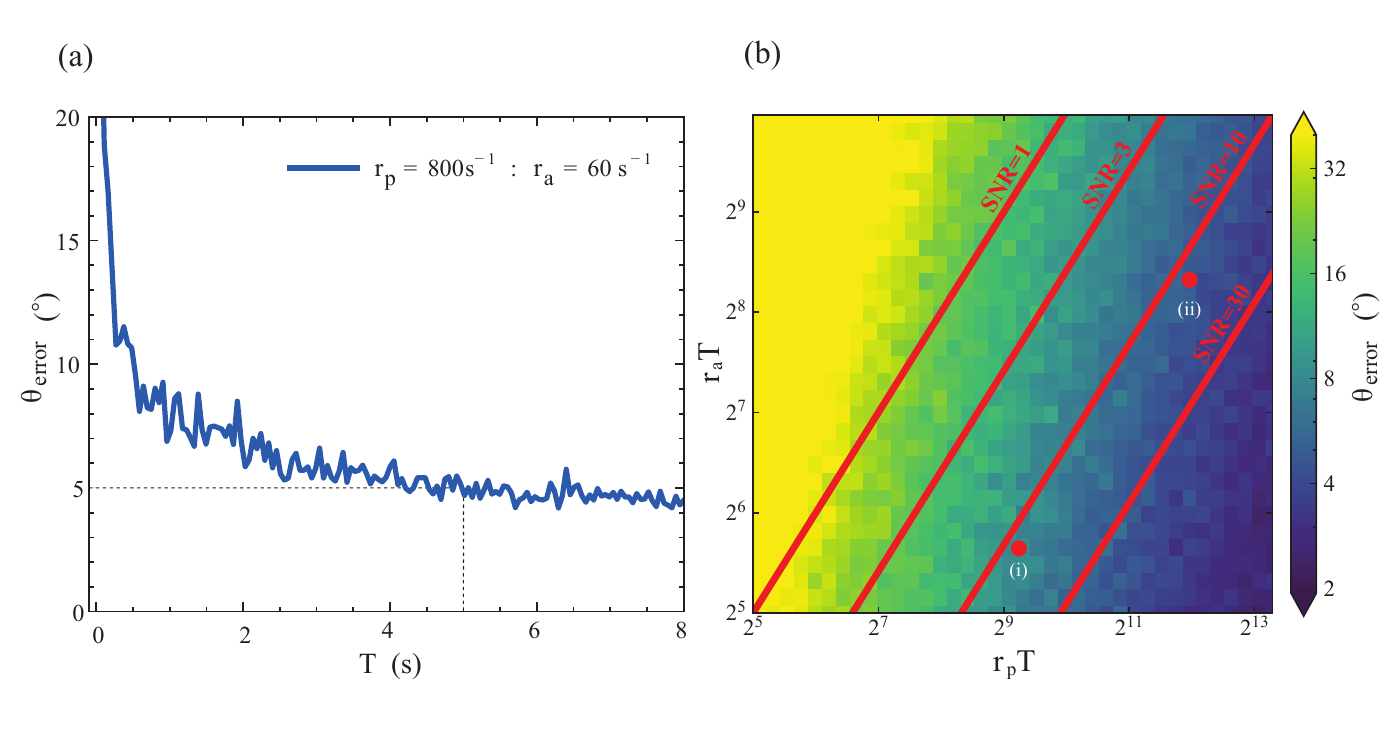}
    \caption{(a) Simulated alignment error as a function of the integration period $T$, assuming Poisson counting statistics for the coincidence detection rate \eqref{eq:13}. (b) Alignment error as a function of the average pair detection level $r_p T$ and the average accidental coincidence count $r_a T$. The lines indicate contours of constant signal-to-noise (coincidence-to-accidental) ratios. The points marked (i) and (ii) correspond to the experimental data rates of the deployed and in-lab fiber spans respectively.}
    \label{fig:6sim}
\end{figure}

To assess the impact of quantization noise and integration time on alignment, we conducted a series of Monte Carlo numerical simulations using different values of $r_p$, $r_a$ and $T$. The simulations employ the same Nelder-Mead algorithm as the experiments and the polarization actuators were modeled as three successive variable retarder waveplates oriented at $0^\circ$, $45^\circ$, and $0^\circ$. After converging to a coincidence minimum, we evaluate the angular deviation between Alice's final polarization state $\vv{a}$ and Bob's target state $\vv{b}$ as $\theta_{\text{error}} = \cos^{-1}(\vv{a}\cdot \vv{b)}$. $\theta_{\text{error}}$ is then averaged over 100 uniformly distributed target states on the Poincar\'e sphere for each data point. Fig.\ref{fig:6sim}(a) shows the simulated alignment error $\theta_{\text{error}}$ as a function of the integration time $T$, for $r_p$ = 800 pairs/s and $r_a$ = 60 pairs/s, which correspond to one set of experimental conditions considered here.  With these data rates we chose five seconds as a compromise between a low polarization alignment error angle and a fast total alignment time.

Fig.~\ref{fig:6sim}(b) shows the average alignment error as a function of $r_pT$ and $r_aT$.  As expected, for a fixed average background count level ($r_a T$), the alignment accuracy improves with increasing average pair count ($r_p T$).  Conversely, for a fixed average pair count, increasing the average background count decreases the alignment accuracy.  The diagonal lines are contours of constant coincidence-to-accidental ratio ($r_p/r_a$), indicating the performance that can be achieved by adjusting the integration period $T$ alone, when the count rates are otherwise fixed.  The two data points, (i) and (ii), within figure~\ref{fig:6sim}(b) mark the experimental conditions used for the deployed fiber loopback configuration (with $T=20$ seconds) and for back-to-back conditions (with $T=5$ seconds), respectively.  Though not shown here, the simulations produced an extinction ratio approaching $r_a/r_p$, provided the method converges.

\section{Polarization Tracking}

\begin{figure}[htbp]
    \centering
    \includegraphics[]{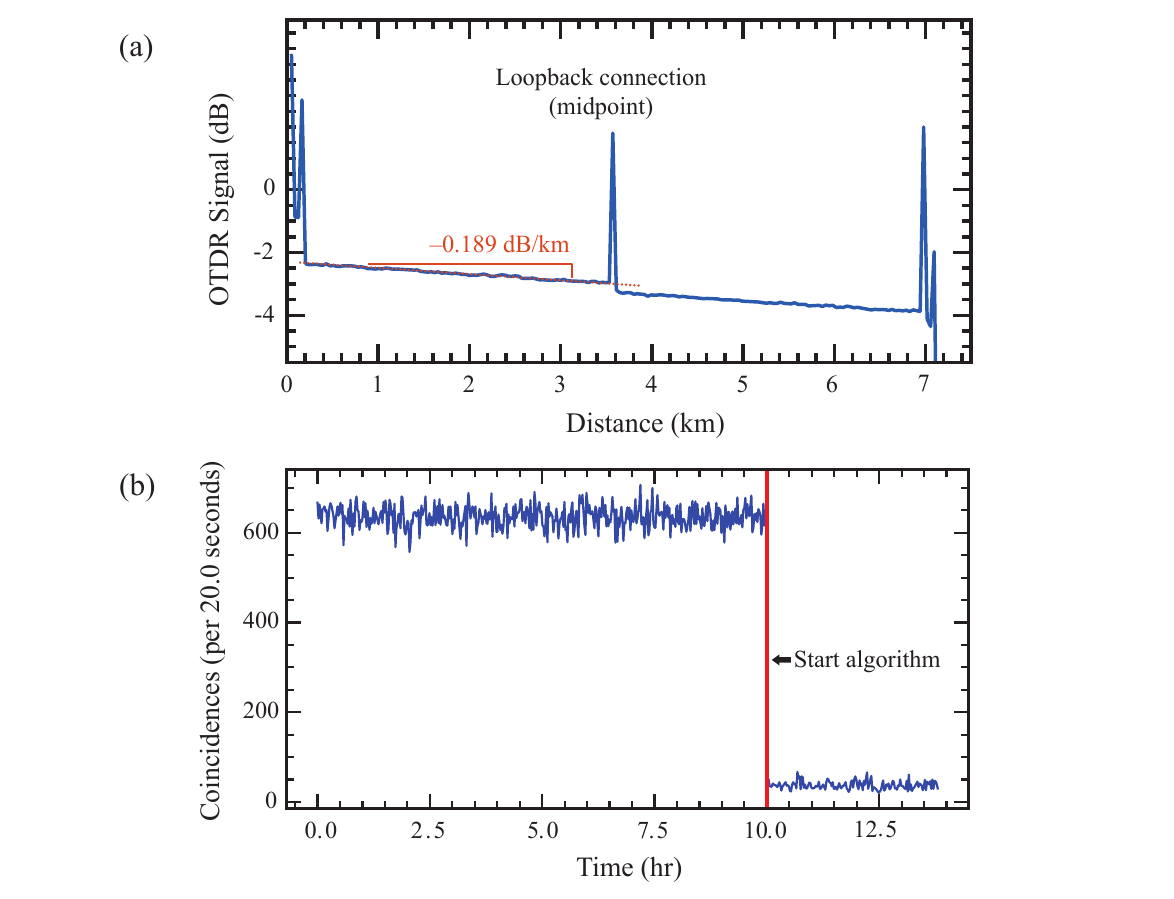}
    \caption{(a) Optical time-domain reflectometry (OTDR) data of the deployed 7.1 km fiber span inserted into Bob's channel, to evaluation polarization tracking method. (b) Measured coincidence rate between Bob's $h_b$ detector and Alice's $h_a$ detector, before and after enabling the alignment algorithm. }
    \label{fig:7}
\end{figure}

To evaluate the capability of the method in a real fiber transmission system, we used a 7.1 km deployed dark fiber link composed of a metropolitan buried SMF-28 singlemode fiber in a loopback configuration.  Fig.~\ref{fig:7}(a) shows an optical time domain reflectometry (OTDR) trace of the fiber span, which has a net insertion loss of 4.2 dB, and a total round-trip delay of 34.6 $\mu$s.  Prior to alignment, we used a classical laser and polarimeter to observe the polarization evolution in this loop over a typical 72 hour period, and observed an average drift rate of $0.1^\circ$ per hour in the interior angle between the initial Stokes vector and the Stokes vector at later times.  When the 7.1 km loop was inserted into Bob's channel, our tracking algorithm was able to quickly align Alice's detector basis and maintain alignment over a 4 hour period, as shown in Fig.~\ref{fig:7}(b).

\begin{figure}[htbp]
    \centering
    \includegraphics[]{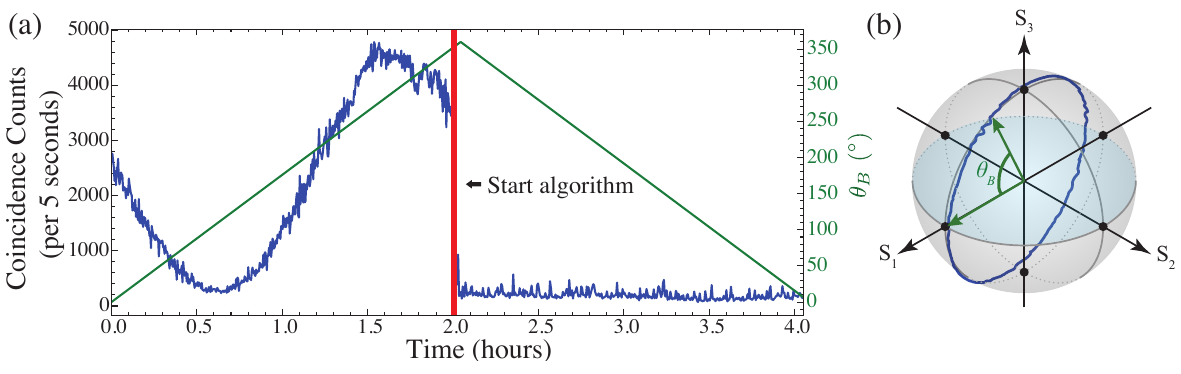}
    \caption{Alice tracking Bob's drifting measurement base. a) Coincidence counts between Alice's and Bob's horizontal detectors when Bob's measurement base drifts at a speed, $\omega_B = \Delta \theta_B/ \Delta t = 176.1^\circ$ per hour. b) An example of a great circle sweep of Bob's measurement base where the rotation angle $\theta_B(t)$ is controlled by the voltage.}
    \label{fig:8}
\end{figure}

Because the drifts experienced in our deployed fiber are slow, we tested the tracking capability of our method by introducing a controlled drift in Bob's measurement basis using the piezoelectric actuator.  Again using a classical laser and polarimeter, the first three piezoelectric stages in Bob's actuator were adjusted such that remaining actuator caused the polarization to trace a great circle, thus providing the maximum amplitude of drift.  The voltage across the fourth actuator controls the rotational angle $\theta_B$ seen in Fig.~\ref{fig:8}b.  This actuator was then programmed with a sawtooth waveform with an voltage amplitude of $V_\pi$ and a slope corresponding to $180^\circ$/hr (shown by the green curve in Fig.~\ref{fig:8}(a)) which is nearly three orders of magnitude faster than what was observed in the 7.1 km deployed fiber.  An additional manual polarization controller then projects this great circle into the nodes of Bob's PBS.  Fig.~\ref{fig:8} shows the great variability in coincidence counts observed during the first two hours when the alignment and tracking algorithm was disabled, and the subsequent minimization of the coincidence rate over a 2 hour period after the algorithm is enabled.

The main bottleneck in the algorithm’s total alignment time and tracking speed is the coincidence count integration time, which determines the rate at which the coincidence rate can be sampled and the polarization setting updated.  As shown in Fig. 6, the integration period required for accurate alignment also depends upon the counting rate, which decreases with the link loss and detection efficiency.

To assess the tracking capability of our system, we extended the Monte Carlo simulations to model the case when the target polarization state $\vv{b}(t)$ steadily traverses a great circle on the Poincar\'e sphere. While the tracking capability also depends upon the noise levels and numerical parameters of the algorithm, for the experimental rates considered here, simulations reveal that the algorithm could maintain alignment provided the target was moving slower than $0.5^\circ$ per integration period.

The timescale of polarization drift in optical fibers depends upon the length of the fiber, as well as the deployed environment (spooled, aerial, buried or subsea).  For the 7.1 km length buried fiber considered here, the drift rate was well below this limit, even for integration periods as long as 20 s.  The forced drifting measurements in Fig.~\ref{fig:8} show successful tracking when the target is drifting $0.25^\circ$ per measurement period.

\section{Conclusion}

In this paper we report a way to use non-local correlations of an entangled state to orient and dynamically track two spatially separated measurement bases.  We present a theory for the coincidence rate for arbitrarily polarized measurement bases, which generalizes the more commonly used result from linearly polarized bases.  The alignment routine uses the Nelder-Mead algorithm to manipulate the polarization controller of one individual to minimize the joint coincidence counts between separate observers. The experimental implementation acquires alignment and minimizes the coincidence rate within approximately 160 seconds.

The alignment strategy is confirmed to work for arbitrary initial orientation of the two observers' measurement bases, and can achieve alignment by minimizing coincidences even when the underlying photon count rates are observed to be statistically unpolarized.  The method relies on the entangled state $\ket{\Psi^-}$, and we show that when the entanglement is broken, the method fails to converge.

In addition to acquiring alignment to a stationary basis, the method is able to dynamically track and stabilize the measurement of entanglement over a 7.1 km deployed fiber loop over several hours of operation, and also under a controlled drifting environment.

Our method relies only on the entangled state, and does not employ classical polarized pilot tones, temporal multiplexers, wavelength filters or additional polarimetry in order to achieve alignment.  This technique could find applications in quantum key distribution, teleportation, entanglement swapping, or locality-violation experiments.  Although these applications may require a series of coordinated sequential measurements between receivers, the polarization alignment and tracking method presented here lays the groundwork for more sophisticated alignment methods demanded by future quantum measurement systems.

\begin{backmatter}

\bmsection{Disclosures}
The authors declare no conflicts of interest.

\bmsection{Data availability}
Data underlying the results presented in this paper are not publicly available at this time but may be obtained from the authors upon reasonable request.

\end{backmatter}

\bibliography{Entangled-Polarization-Control-2022}

\begin{thebibliography}{10}
\newcommand{\enquote}[1]{``#1''}

\bibitem{Nielsen2012}
M.~A. Nielsen and I.~L. Chuang, \emph{{Quantum Computation and Quantum
  Information: 10th Anniversary Edition}} (Cambridge University Press, 2012).

\bibitem{Giustina2015}
M.~Giustina, M.~A.~M. Versteegh, S.~Wengerowsky, J.~Handsteiner, A.~Hochrainer,
  K.~Phelan, F.~Steinlechner, J.~Kofler, J.-A. Larsson, C.~Abell\'an, W.~Amaya,
  V.~Pruneri, M.~W. Mitchell, J.~Beyer, T.~Gerrits, A.~E. Lita, L.~K. Shalm,
  S.~W. Nam, T.~Scheidl, R.~Ursin, B.~Wittmann, and A.~Zeilinger,
  \enquote{{Significant-Loophole-Free Test of Bell's Theorem with Entangled
  Photons},} {\protect\JournalTitle{Phys. Rev. Lett.}} \textbf{115}, 250401
  (2015).

\bibitem{Bennett1992}
C.~H. Bennett, G.~Brassard, and N.~D. Mermin, \enquote{{Quantum cryptography
  without Bell's theorem},} {\protect\JournalTitle{Phys. Rev. Lett.}}
  \textbf{68}, 557--559 (1992).

\bibitem{Jin2015}
R.-B. Jin, M.~Takeoka, U.~Takagi, R.~Shimizu, and M.~Sasaki, \enquote{{Highly
  efficient entanglement swapping and teleportation at telecom wavelength},}
  {\protect\JournalTitle{Sci. Rep.}} \textbf{5}, 9333 (2015).

\bibitem{Schultz1979}
P.~C. Schultz, \enquote{{Progress in optical waveguide process and materials},}
  {\protect\JournalTitle{Appl. Opt.}} \textbf{18}, 3684--3693 (1979).

\bibitem{Miya1979}
T.~Miya, Y.~Terunuma, T.~Hosaka, and T.~Miyashita, \enquote{{Ultimate low-loss
  single-mode fibre at 1.55 $\mu$m},} {\protect\JournalTitle{Electron. Lett.}}
  \textbf{15}, 106--108 (1979).

\bibitem{Dynes2009}
J.~F. Dynes, H.~Takesue, Z.~L. Yuan, A.~W. Sharpe, K.~Harada, T.~Honjo,
  H.~Kamada, O.~Tadanaga, Y.~Nishida, M.~Asobe, and A.~J. Shields,
  \enquote{Efficient entanglement distribution over 200 kilometers,}
  {\protect\JournalTitle{Opt. Express}} \textbf{17}, 11440--11449 (2009).

\bibitem{Inagaki2013}
T.~Inagaki, N.~Matsuda, O.~Tadanaga, M.~Asobe, and H.~Takesue,
  \enquote{{Entanglement distribution over 300 km of fiber},}
  {\protect\JournalTitle{Opt. Express}} \textbf{21}, 23241--23249 (2013).

\bibitem{Hong1985}
C.~K. Hong and L.~Mandel, \enquote{{Theory of parametric frequency down
  conversion of light},} {\protect\JournalTitle{Phys. Rev. A}} \textbf{31},
  2409--2418 (1985).

\bibitem{Burnham1970}
D.~C. Burnham and D.~L. Weinberg, \enquote{{Observation of Simultaneity in
  Parametric Production of Optical Photon Pairs},} {\protect\JournalTitle{Phys.
  Rev. Lett.}} \textbf{25}, 84--87 (1970).

\bibitem{Zhang2021}
C.~Zhang, Y.-F. Huang, B.-H. Liu, C.-F. Li, and G.-C. Guo,
  \enquote{{Spontaneous Parametric Down-Conversion Sources for Multiphoton
  Experiments},} {\protect\JournalTitle{Adv. Quantum Technol.}} \textbf{4},
  2000132 (2021).

\bibitem{Imai1988}
T.~Imai and T.~Matsumoto, \enquote{{Polarization fluctuations in a single-mode
  optical fiber},} {\protect\JournalTitle{Journal of Lightwave Technology}}
  \textbf{6}, 1366--1375 (1988).

\bibitem{Namihira1989}
Y.~Namihira and H.~Wakabayashi, \enquote{{Real-time measurements of
  polarization fluctuations in an optical fiber submarine cable in a deep-sea
  trial using electrooptic LiNbO$_3$ device},} {\protect\JournalTitle{Journal
  of Lightwave Technology}} \textbf{7}, 1201--1206 (1989).

\bibitem{Karlsson2000}
M.~Karlsson, J.~Brentel, and P.~A. Andrekson, \enquote{Long-term measurement of
  pmd and polarization drift in installed fibers,}
  {\protect\JournalTitle{Journal of Lightwave Technology}} \textbf{18},
  941--951 (2000).

\bibitem{Waddy2005}
D.~S. Waddy, L.~Chen, and X.~Bao, \enquote{{Polarization effects in aerial
  fibers},} {\protect\JournalTitle{Optical Fiber Technology}} \textbf{11},
  1--19 (2005).

\bibitem{Ding2017a}
Y.-Y. Ding, H.~Chen, S.~Wang, D.-Y. He, Z.-Q. Yin, W.~Chen, Z.~Zhou, G.-C. Guo,
  and Z.-F. Han, \enquote{{Polarization variations in installed fibers and
  their influence on quantum key distribution systems},}
  {\protect\JournalTitle{Opt. Express}} \textbf{25}, 27923--27936 (2017).

\bibitem{Noe1988}
R.~No{\'{e}}, H.~Heidrich, and D.~Hoffmann, \enquote{{Endless Polarization
  Control Systems for Coherent Optics},} {\protect\JournalTitle{Journal of
  Lightwave Technology}} \textbf{6}, 1199--1208 (1988).

\bibitem{Walker1990}
N.~G. Walker and G.~R. Walker, \enquote{{Polarization Control for Coherent
  Communications},} {\protect\JournalTitle{Journal of Lightwave Technology}}
  \textbf{8}, 438--458 (1990).

\bibitem{Shimizu1991}
H.~Shimizu, S.~Yamazaki, T.~Ono, and K.~Emura, \enquote{{Highly practical fiber
  squeezer polarization controller},} {\protect\JournalTitle{J. Lightwave
  Technol.}} \textbf{9}, 1217--1224 (1991).

\bibitem{Heismann1994}
F.~Heismann, \enquote{{Analysis of a Reset-Free Polarization Controller for
  Fast Automatic Polarization Stabilization in Fiber-optic Transmission
  Systems},} {\protect\JournalTitle{Journal of Lightwave Technology}}
  \textbf{12}, 690--699 (1994).

\bibitem{Martinelli2003}
M.~Martinelli and R.~A. Chipman, \enquote{{Endless Polarization Control
  Algorithm Using Adjustable Linear Retarders With Fixed Axes},}
  {\protect\JournalTitle{Journal of Lightwave Technology}} \textbf{21},
  2089--2096 (2003).

\bibitem{Martinelli2006}
M.~Martinelli, P.~Martelli, and S.~M. Pietralunga, \enquote{{Polarization
  stabilization in optical communications systems},}
  {\protect\JournalTitle{Journal of Lightwave Technology}} \textbf{24},
  4172--4183 (2006).

\bibitem{Xavier2008}
G.~B. Xavier, G.~Vilela~de Faria, G.~P. Tempor{\~{a}}o, J.~P. von~der Weid,
  z.~Peng, J.~Zhang, D.~Yang, W.-b. Gao, H.-x. Ma, H.~Yin, H.-p. Zeng, T.~Yang
  X-B~Wang, J.~Waldeb{\"{a}}ck, M.~Tengner, D.~Ljunggren, and A.~Karlsson,
  \enquote{{Full polarization control for fiber optical quantum communication
  systems using polarization encoding},} {\protect\JournalTitle{Optics
  Express}} \textbf{16}, 1867--1873 (2008).

\bibitem{Xavier2009}
G.~B. Xavier, N.~Walenta, G.~V. de~Faria, G.~P. Tempor{\~{a}}o, N.~Gisin,
  H.~Zbinden, and J.~P. von~der Weid, \enquote{{Experimental polarization
  encoded quantum key distribution over optical fibres with real-time
  continuous birefringence compensation},} {\protect\JournalTitle{New J.
  Phys.}} \textbf{11}, 045015 (2009).

\bibitem{Chen2009}
J.~Chen, G.~Wu, L.~Xu, X.~Gu, E.~Wu, and H.~Zeng, \enquote{{Stable quantum key
  distribution with active polarization control based on time-division
  multiplexing},} {\protect\JournalTitle{New J. Phys.}} \textbf{11}, 065004
  (2009).

\bibitem{Li2018}
D.-D. Li, S.~Gao, G.-C. Li, L.~Xue, L.-W. Wang, C.-B. Lu, Y.~Xiang, Z.-Y. Zhao,
  L.-C. Yan, Z.-Y. Chen, G.~Yu, and J.-H. Liu, \enquote{{Field implementation
  of long-distance quantum key distribution over aerial fiber with fast
  polarization feedback},} {\protect\JournalTitle{Opt. Express}} \textbf{26},
  22793--22800 (2018).

\bibitem{Wang2009}
S.~X. Wang and G.~S. Kanter, \enquote{{Robust Multiwavelength All-Fiber Source
  of Polarization-Entangled Photons With Built-In Analyzer Alignment Signal},}
  {\protect\JournalTitle{IEEE J. Sel. Top. Quantum Electron.}} \textbf{15},
  1733--1740 (2009).

\bibitem{Chung2022}
J.~Chung, E.~M. Eastman, G.~S. Kanter, K.~Kapoor, N.~Lauk, C.~H. Pe\~na, R.~K.
  Plunkett, N.~Sinclair, J.~M. Thomas, R.~Valivarthi, S.~Xie, R.~Kettimuthu,
  P.~Kumar, P.~Spentzouris, and M.~Spiropulu, \enquote{{Design and
  Implementation of the Illinois Express Quantum Metropolitan Area Network},}
  {\protect\JournalTitle{IEEE Trans. Quantum Eng.}} \textbf{3}, 1--20 (2022).

\bibitem{Ding2017b}
Y.-Y. Ding, W.~Chen, H.~Chen, C.~Wang, Y.-P. li, S.~Wang, Z.-Q. Yin, G.-C. Guo,
  and Z.-F. Han, \enquote{{Polarization-basis tracking scheme for quantum key
  distribution using revealed sifted key bits},} {\protect\JournalTitle{Opt.
  Lett.}} \textbf{42}, 1023--1026 (2017).

\bibitem{Agnesi2020}
C.~Agnesi, M.~Avesani, L.~Calderaro, A.~Stanco, G.~Foletto, M.~Zahidy,
  A.~Scriminich, F.~Vedovato, G.~Vallone, and P.~Villoresi, \enquote{{Simple
  quantum key distribution with qubit-based synchronization and a
  self-compensating polarization encoder},} {\protect\JournalTitle{Optica}}
  \textbf{7}, 284--290 (2020).

\bibitem{Shi2021}
Y.~Shi, H.~S. Poh, A.~Ling, and C.~Kurtsiefer, \enquote{{Fibre polarisation
  state compensation in entanglement-based quantum key distribution},}
  {\protect\JournalTitle{Opt. Express}} \textbf{29}, 37075--37080 (2021).

\bibitem{Cortes2022}
C.~L. Cortes, P.~Lefebvre, N.~Lauk, M.~J. Davis, N.~Sinclair, S.~K. Gray, and
  D.~Oblak, \enquote{{Sample-efficient adaptive calibration of quantum networks
  using Bayesian optimization},} {\protect\JournalTitle{Physical Review
  Applied}} \textbf{17}, 034067 (2022).

\bibitem{Fiorentino2007}
M.~Fiorentino, S.~M. Spillane, R.~G. Beausoleil, T.~D. Roberts, P.~Battle, and
  M.~W. Munro, \enquote{{Spontaneous parametric down-conversion in periodically
  poled KTP waveguides and bulk crystals},} {\protect\JournalTitle{Opt.
  Express}} \textbf{15}, 7479--7488 (2007).

\bibitem{Couteau2018}
C.~Couteau, \enquote{{Spontaneous parametric down-conversion},}
  {\protect\JournalTitle{Contemporary Physics}} \textbf{59}, 291--304 (2018).

\bibitem{Kuklewicz2004}
C.~E. Kuklewicz, M.~Fiorentino, G.~Messin, F.~N.~C. Wong, and J.~H. Shapiro,
  \enquote{{High-flux source of polarization-entangled photons from a
  periodically poled ${\mathrm{KTiOPO}}_{4}$ parametric down-converter},}
  {\protect\JournalTitle{Phys. Rev. A}} \textbf{69}, 013807 (2004).

\bibitem{Takeoka2015}
M.~Takeoka, R.-B. Jin, and M.~Sasaki, \enquote{{Full analysis of multi-photon
  pair effects in spontaneous parametric down conversion based photonic quantum
  information processing},} {\protect\JournalTitle{New J. Phys.}} \textbf{17},
  043030 (2015).

\bibitem{Brewster2021}
R.~A. Brewster, G.~Baumgartner, and Y.~K. Chembo, \enquote{{Quantum analysis of
  polarization entanglement degradation induced by multiple-photon-pair
  generation},} {\protect\JournalTitle{Phys. Rev. A}} \textbf{104}, 022411
  (2021).

\bibitem{Gordon2000}
J.~P. Gordon and H.~Kogelnik, \enquote{{PMD fundamentals: Polarization mode
  dispersion in optical fibers},} {\protect\JournalTitle{Proc. Natl. Acad. Sci.
  U.S.A.}} \textbf{97}, 4541--4550 (2000).

\bibitem{Jones2018}
D.~E. Jones, B.~T. Kirby, and M.~Brodsky, \enquote{{Tuning quantum channels to
  maximize polarization entanglement for telecom photon pairs},}
  {\protect\JournalTitle{npj Quantum Inf.}} \textbf{4}, 58 (2018).

\bibitem{Nelder1965}
J.~A. Nelder and R.~Mead, \enquote{{A Simplex Method for Function
  Minimization},} {\protect\JournalTitle{The Computer Journal}} \textbf{7},
  308--313 (1965).

\bibitem{Virtanen2020}
P.~Virtanen, R.~Gommers, T.~E. Oliphant, M.~Haberland, T.~Reddy, D.~Cournapeau,
  E.~Burovski, P.~Peterson, W.~Weckesser, J.~Bright, S.~J. van~der Walt,
  M.~Brett, J.~Wilson, K.~J. Millman, N.~Mayorov, A.~R.~J. Nelson, E.~Jones,
  R.~Kern, E.~Larson, C.~J. Carey, İlhan Polat, Y.~Feng, E.~W. Moore,
  J.~VanderPlas, D.~Laxalde, J.~Perktold, R.~Cimrman, I.~Henriksen, E.~A.
  Quintero, C.~R. Harris, A.~M. Archibald, A.~H. Ribeiro, F.~Pedregosa, P.~van
  Mulbregt, A.~Vijaykumar, A.~P. Bardelli, A.~Rothberg, A.~Hilboll,
  A.~Kloeckner, A.~Scopatz, A.~Lee, A.~Rokem, C.~N. Woods, C.~Fulton,
  C.~Masson, C.~Häggström, C.~Fitzgerald, D.~A. Nicholson, D.~R. Hagen, D.~V.
  Pasechnik, E.~Olivetti, E.~Martin, E.~Wieser, F.~Silva, F.~Lenders,
  F.~Wilhelm, G.~Young, G.~A. Price, G.-L. Ingold, G.~E. Allen, G.~R. Lee,
  H.~Audren, I.~Probst, J.~P. Dietrich, J.~Silterra, J.~T. Webber, J.~Slavič,
  J.~Nothman, J.~Buchner, J.~Kulick, J.~L. Schönberger, J.~V.
  de~Miranda~Cardoso, J.~Reimer, J.~Harrington, J.~L.~C. Rodríguez,
  J.~Nunez-Iglesias, J.~Kuczynski, K.~Tritz, M.~Thoma, M.~Newville,
  M.~Kümmerer, M.~Bolingbroke, M.~Tartre, M.~Pak, N.~J. Smith, N.~Nowaczyk,
  N.~Shebanov, O.~Pavlyk, P.~A. Brodtkorb, P.~Lee, R.~T. McGibbon,
  R.~Feldbauer, S.~Lewis, S.~Tygier, S.~Sievert, S.~Vigna, S.~Peterson,
  S.~More, T.~Pudlik, and T.~Oshima, \enquote{{SciPy 1.0: fundamental
  algorithms for scientific computing in Python},}
  {\protect\JournalTitle{Nature Methods}} \textbf{17}, 261--272 (2020).

\end{thebibliography}

\end{document}